%% file: eux_distance_laws_18.tex
\documentclass[aps,prb,twocolumn,superscriptaddress,showpacs,byrevtex]{revtex4}

\usepackage[latin1]{inputenc}
\usepackage{graphicx,textcomp}
\usepackage{bm,amsmath}

\usepackage{color}

\begin{document}


\title{Exchange interactions in europium monochalcogenide magnetic
semiconductors and their dependence on hydrostatic strain}


\author{W.~S\"ollinger}
\author{W.~Heiss}
\author{R.~T.~Lechner}
\affiliation{Institute of Semiconductor and Solid State Physics, Johannes Kepler
University, Altenbergerstrasse 69, A-4040 Linz, Austria}
\author{K.~Rumpf}
\author{P.~Granitzer}
\author{H.~Krenn}
\affiliation{Institute of Physics, Experimental Physics Div.,
Karl-Franzens-University, Universit\"atsplatz 5, A-8010 Graz, Austria}
\author{G.~Springholz}
\thanks{To whom correspondence should be addressed. E-mail: gunther.springholz@jku.at}
\affiliation{Institute of Semiconductor and Solid State Physics, Johannes Kepler
University, Altenbergerstrasse 69, A-4040 Linz, Austria}


\date{\today}

\begin{abstract}
The classical Heisenberg model is applied in a Monte Carlo study
to investigate the distance dependence of the indirect nearest
neighbor (NN) exchange and next-nearest neighbor (NNN)
superexchange interaction in EuO, EuS, EuSe and EuTe. For this
purpose, first, the dependence of the magnetic ordering
temperature, i.e., Curie, respectively, N\'eel temperature for
ferromagnetic and antiferromagnetic ordering on the exchange
constants was determined. This was then employed for the analysis
of experimental data of hydrostatic pressure experiments. It is
shown that all experimental findings, i.e., the strong increase of
the critical temperatures, as well as the transition from
antiferromagnetic to ferromagnetic ordering for EuTe and EuSe with
decreasing lattice parameter is well described by the magnetic
Gr\"uneisen law, in which the exchange constants depend on the
interatomic distances of the Eu ions in the form of a power law.
According to these calculations, the indirect NN exchange is
characterized by a Gr\"uneisen exponent of approximately 20 and
the NNN superexchange by an exponent of about 10 for all four
europium monochalcogenides. The latter agrees with Bloch's
empirical 10/3 law for the volume dependence of superexchange
interactions in insulating magnetic materials. The Monte Carlo
calculations also yield significantly revised exchange constants
for unstrained bulk material because spin fluctuations at non-zero
temperatures are taken into account. The strong increase of the
exchange constants with decreasing lattice parameter provides room
for increasing the Curie temperatures in strained epitaxial
structures, which is important for device applications.

\end{abstract}


\maketitle



\section{Introduction\label{sec:intro}}

The europium monochalcogenides (Eu$X$, with $X = \mbox{O}$, S, Se
or Te) are wide band gap magnetic semiconductors with cubic
rocksalt crystal structure and increasing lattice constant as $X$
changes from O to Te\cite{zinn76,wacht79,mauge86}.
They are considered to be model substances
for Heisenberg magnets with spin ordering dominated by
indirect nearest neighbor (NN) exchange $J_1$ and next-nearest
neighbor (NNN) superexchange  $J_2$ acting between the $S = 7/2$
localized magnetic moments of the Eu\textsuperscript{2+} ions with
half filled $4f$ shells\cite{kasuy70}. Depending on the sign and magnitude of
the exchange integrals $J_1$ and $J_2$, the Eu$X$s exhibit
different magnetic phases below the critical ordering
temperature\cite{seehr88}.
EuO\cite{matth61} and EuS\cite{mcgui62} are ferromagnets and
EuTe\cite{olive72} is an antiferromagnet. EuSe
is at the borderline between ferromagnetic and antiferromagnetic
ordering. Thus, it shows metamagnetic behavior\cite{gries71,fukum85,lechn05},
which is influenced by additional contributions from dipolar interactions
and crystalline field anisotropies.

The Eu$X$s show several outstanding  properties, which makes them
an interesting class of materials, both academically and for
device applications. In external magnetic fields they exhibit a
giant spin-splitting of the conduction band and, consequently,
extraordinary large magneto-optical effects. EuSe shows the
largest effective $g$-factor\cite{kirch04} of up to 18\,000 and
EuTe the largest magnetic field induced energy shifts of the
interband transitions\cite{heiss01} observed in semiconducting
materials. Potential applications are spin-filter devices based on
EuO\cite{steen02,santo04,santo08},
EuS~\cite{moode88,hao90,filip02,lecla02,smits04,trbov05,nagah07,ren07}
or EuSe\cite{moode93} tunnel junctions, which provide
spin-polarized electrons due to different barrier heights for
electrons in different spin states. Also, a huge Faraday rotation
is observed in Eu$X$s\cite{busch70,schoe77,sueka83,hori94} due to
the spin-splitting of the bands, which results in different
refractive indices for left and right circular polarized light.
Therefore, EuS/EuF$_2$ and EuSe  films have been used for high
resolution magneto-optical imaging of the flux distribution in
superconductors\cite{kobli95}. Recent work has also demonstrated
that EuO can be epitaxially grown on silicon
\cite{letti03,schme07} and GaN\cite{schme07}, which opens new
possibilities for device realization. Since the Curie temperature
of the Eu$X$s can be drastically enhanced by
doping\cite{konno96,ott06,schme07}, EuO might even become a
candidate for practical spintronic device applications.

Introducing strain, either omniaxially through hydrostatic
pressure\cite{mcwha66,sriva68,klein76,klein77,moser79,fujiw82,sauer83,
hihar85,tisse87,moser88,abdel90,gonch97,ishiz97,gonch98} or
biaxially through epitaxial strain
\cite{spring93a,frank94,stach99,kepa03,lechn05,schie08} leads to
drastic changes in the ordering temperatures in the Eu$X$
compounds and in some cases even to transitions to different kinds
of magnetic ordering. For EuO, hydrostatic pressure was found to
increase  the ferromagnetic ordering temperature $T_C$ from 69 to
above 200\,K\cite{abdel90} and for EuS from 16 to almost
180\,K\cite{gonch98}. Metamagnetic EuSe is transformed to a stable
ferromagnet already at moderate hydrostatic pressures above
0.5\,GPa\cite{fujiw82} and at higher pressures $T_C$ increases
from 4.7 to 70\,K at 15\,GPa\cite{gonch98}. EuTe remains
antiferromagnetic up to 9\,GPa with nearly constant N\'eel
temperature $T_N \approx 10$\,K, but then becomes ferromagnetic
with a $T_C$ increasing up to 28\,K\ when reaching
17\,GPa\cite{gonch97}.

The variations in the magnetic properties of the Eu$X$ compounds
are obviously related to the dependence of the exchange integrals
$J_1$ and $J_2$ on the interatomic distances in the crystal
lattice. Already by early theoretical work, the basic trend of the
Eu$X$ compounds from antiferromagnetic (EuTe) to ferromagnetic
ordering (EuS and EuO) was attributed to a strong increase of the
ferromagnetic NN exchange $J_1$ with decreasing lattice constant
from $a_0 = 6.598$\,\AA\ for EuTe to 5.144\,\AA\ for EuO. For the
latter, the magnetic ordering is thus dominated by the positive NN
exchange $J_1$, whereas for antiferromagnetic EuTe the negative
NNN exchange $J_2$ dominates. Application of hydrostatic pressures
$p$ up to 20\,GPa, produces similar changes in the lattice
parameter of up to 8\,\% compared to the normal bulk values. As a
result, large changes in the ordering temperatures are induced as
well
\cite{mcwha66,sriva68,klein76,klein77,moser79,fujiw82,sauer83,
hihar85,tisse87,moser88,abdel90,gonch97,ishiz97,gonch98}.

To derive the dependence of the exchange integrals $J_1$ and $J_2$
on the interatomic distances from hydrostatic pressure
experiments, previous works have employed the mean field
approximation (MFA) for analysis\cite{gonch97,gonch98,gonch00}.
Based on the observation that the N\'eel temperature $T_N$ of
antiferromagnetic EuTe does not change appreciably under applied
pressure, it was reasoned that the NNN exchange $J_2$ is constant
in all Eu$X$ compounds. Therefore, the changes in the magnetic
properties were attributed solely to changes in the NN exchange
$J_1$ and, from the simple mean field expressions, a distance
dependence of $J_1(a)$ was deduced from the observed changes of
$T_C(p)$. However, it is well known that the mean field
approximation is exact only at zero temperature, i.e., for
prediction of the ground state of the system. In particular, the
mean field model vastly overestimates the magnetic ordering
temperatures due to neglection of spin fluctuations at finite
temperatures. As a result, the mean field approximation not only
predicts false critical exponents at the phase transition but also
much underrated exchange constants.

In the present work, we have employed the Monte Carlo method (MC)
to calculate the magnetic phase diagrams and ordering temperatures
of the Eu$X$ compounds as a function of the exchange interactions.
The MC method takes the mutual interactions between all spins into
account and allows for spin fluctuations at $T$ above zero. Using
finite size scaling techniques, MC correctly predicts the
transition temperatures and the behavior of the order parameters
at criticality for a given model Hamiltonian\cite{landa00}.
Moreover, due to the spin fluctuations near the transition
temperature in the MC calculations, the N\'eel point of
antiferromagnetic ordering as in EuTe is found to depend on
on the exchange constants $J_1$ and $J_2$\cite{scott02}, in contrast
to the mean field approximation, where the N\'eel point depends on $J_2$ only.
Thus,the basic assumption of the previous analyses does not hold
\cite{gonch97,gonch98,gonch00}. The experimental data for Eu$X$s
under hydrostatic pressure is therefore reexamined by Monte Carlo
calculations and by considering magnetic Gr\"uneisen power laws
\cite{bloch66} for the distance dependence of the NN and NNN
exchange constants.
We show that
for the whole family of Eu$X$ compounds, the distance dependence
of the exchange interactions $J_1(r_1)$ and $J_2(r_2)$ can be
consistently described by unique Gr\"uneisen exponents of $n_1
\approx 20$ and $n_2 \approx 10$ by which the whole set of
experimental data of the Eu$X$ compounds under hydrostatic
pressure can well be explained. The obtained exponent of $n_2
\approx 10$ for $J_2$ is also consistent with Bloch's empirical
10/3 law \cite{bloch66} for the volume dependence of the
superexchange interaction $J \sim V^{-10/3}$ observed for a wide
variety of insulating magnetic material systems.

The paper is organized as follows: In Sec.~\ref{sec:computational}
we present the model Hamiltonian and briefly discuss the technical
details of the Monte Carlo calculations. In Sec.~\ref{sec:eute},
the method is applied to bulk EuTe under ambient pressure,
demonstrating that the whole magnetic phase diagram can be well
described and that the exchange constants $J_1$ and $J_2$ obtained
from Monte Carlo calculations strongly differ from literature
values derived by the mean field approximation. In
Sec.~\ref{sec:distance_eute}, the method for determination of the
distance dependence of the exchange constants is described and
applied to EuTe. Due to the pressure induced transition between
antiferromagnetic and ferromagnetic ordering, accurate
dependencies for both $J_1(r_1)$ and $J_2(r_2)$ are determined.
The approach is then extended to EuO, EuS and EuSe in
Secs.~\ref{sec:euo_eus} and~\ref{sec:euse}, revealing that the
same functional behavior, i.e., the same Gr\"uneisen exponents
provide an excellent description of the experimental data for all
Eu$X$ compounds. In Sec.~\ref{sec:discussion} the results are
compared in detail and the applicability of other types of
functional dependence of $J_1(r_1)$ and $J_2(r_2)$  discussed.

\section{Details of the calculation\label{sec:computational}}

For the calculation of the magnetic properties of the Eu$X$s we employed
the classical Heisenberg model with nearest and next nearest
neighbor exchange interaction taken into account. The corresponding model
Hamiltonian reads as
\begin{equation} \label{eqn:hamiltonian}
 \mathcal H = -\sum_{i \neq j} J_{ij} \bm S_i\bm S_j
    - g \mu_B \bm H \sum_i \bm S_i\quad,
\end{equation}
where $\bm H$ denotes the external magnetic field and $J_{ij}+J_{ji}$ is
the total exchange interaction between two spins located at lattice sites $i$
and $j$ and
\begin{equation}
 J_{ij} = \left\{ \begin{array}{r@{\quad:\quad}l}
                  J_1 & i\mbox{ is NN of }j \\
                  J_2 & i\mbox{ is NNN of }j \\
                  0   & \mbox{else} \end{array} \right.\quad.
\end{equation}
In the Monte Carlo calculations we considered rhombohedral
\textit{fcc} clusters of classical spins, where all cluster
boundaries are (111) lattice planes. This is a convenient choice
of geometry, since antiferromagnetic and ferrimagnetic ordering in
EuSe and EuTe is comprised of ferromagnetic (111) planes and
epitaxial EuTe and EuSe samples are usually grown in (111)
orientation\cite{sprin93,heiss01,kirch04,lechn05}. The
choice of geometry has, however, no influence on the results of our
calculations. Clusters of up to 32\textsuperscript{3} spins with
periodic boundary conditions were considered. During a single
Monte Carlo step, random orientations are generated for every
single spin, which are then accepted or rejected according to the
Metropolis criterion\cite{metro53}. Observeables like the total
energy, the overall magnetization, the magnetization in the
direction of the external field, the transverse magnetization and
the corresponding staggered magnetic moments are computed after
every Monte Carlo step. Simulations were performed with up to
$N=10^5$ iterations and additional Monte Carlo steps for
equilibration at the beginning of every run. For a single
simulation the temperature $T$, the external magnetic field $H$
and the number of spins are constant.

The expectation value for the total energy is given by
\begin{equation}\label{eqn:h_ave}
\langle \mathcal H \rangle \approx \frac{\sum_{i=1}^N \mathcal H_i\exp(-\beta
\mathcal H_i)}{\sum_{i=1}^N
\exp(-\beta \mathcal H_i)},
\end{equation}
where $\beta = 1/k_B\,T$. The Metropolis algorithm causes the total
energy to be distributed according to Boltzmann's law. Hence, the
expectation values for the total energy and for the magnetic
moment become arithmetic mean values in the simulation
\begin{equation}
\left\langle \mathcal H \right\rangle = \frac{1}{N}\sum_{i=1}^N \mathcal H_i
\end{equation}
\begin{equation}
\left\langle \bm M \right\rangle = \frac{1}{N}\sum_{i=1}^N \bm M_i
\end{equation}

To determine the critical ordering temperatures the fourth-order cumulant of
the corresponding order parameter is used, which for ferromagnetic ordering
at zero external field is defined as\cite{landa00}
\begin{equation}
 U = 1-\frac{\langle M^4 \rangle}{3 \langle M^2 \rangle^2}\quad,
\end{equation}
where $\langle M^2 \rangle$ and $\langle M^4 \rangle$ denote the
second and fourth order moments of the probability distribution of
the magnetization. The fourth order cumulants show universal
values at the critical temperature. Thus, during simulation $U$ is
generated as a function of temperature and recorded for various
cluster sizes. The different $U(T)$ curves cross in a single point
at the critical temperature, as shown in detail in
Sec.~\ref{sec:eute}. The transition temperatures obtained from our
calculations were compared to theoretical predictions from high-temperature
series expansions\cite{ritch72}. For
ferromagnetic nearest neighbor exchange ($J_1 > 0$) and no
next-nearest neighbor exchange ($J_2=0$) between classical spins
in the \textit{fcc} lattice, we obtained a critical temperature
$T_C$ defined by $2\,J_1\,S^2/(k_B\,T_C) = 0.3149\pm0.0008$ in
$2\times10^5$ Monte Carlo steps. The prefactor of 2 stems from the
fact, that according to the definition of
Eqn.~(\ref{eqn:hamiltonian}) the exchange interaction between
pairs of spins is always added twice. Despite the simplicity of
our approach compared to other sophisticated Monte Carlo
routines\cite{pecza91,chen93}, our result is in excellent
agreement the theoretically predicted and generally accepted value
of Ritchie and Fisher\cite{ritch72} of $0.3147\pm0.0001$ for this
relation.

If the NNN exchange interaction is antiferromagnetic and $|J_2| >
J_1$, the simulation generates a classical N\'eel state with eight
ferromagnetically ordered sublattices as the ground
state\cite{nowot89}. This is a consequence of the
antiferromagnetic ordering degenerating into the four equivalent
(111) directions\cite{kepa03} in the simulation. In this case,
four pairs of antiferromagnetically aligned sublattices can rotate
freely and the fourth order cumulants are defined as\cite{nowot89}
\begin{equation} \label{eqn:U_st}
U^{st}=\frac{5}{2}-\frac{3}{2}\frac{\langle(M^{st})^4\rangle}{\langle(M^{ st }
)^2\rangle^2}
\end{equation}
for zero field and
\begin{equation} \label{eqn:U_trans_st}
U_{\bot}^{st}=2-\frac{\langle(M_{\bot}^{st})^4\rangle}{\langle(M_{\bot}^{ st }
)^2\rangle^2}\quad,
\end{equation}
for nonzero external field. Here, $M^{st}$ and $M_{\bot}^{st}$ denote the
staggered (transverse) magnetization, with transverse referring to the component
of the magnetization perpendicular to the external field.
\begin{equation} \label{eqn:M_st}
M^{st} = \sum_{i=1}^8 |\bm M^{(i)}|
\end{equation}
\begin{equation} \label{eqn:M_trans_st}
M_\bot^{st} = \sum_{i=1}^8 |\bm M_\bot^{(i)}|
\end{equation}
Eqns.~(\ref{eqn:M_st}) and~(\ref{eqn:M_trans_st}) are sums of the
absolute values of the (transverse) magnetization over the eight
possible sublattices. Due to additional anisotropies and/or
dipolar couplings, this degenerate AFM state is not observed
experimentally, but only domains with completely ferromagnetically
ordered Eu (111) planes, which is one possible case in our MC
simulations.
Here it should be also noted that in a quantum mechanical
treatment it has been shown\cite{nolti86} that a classical N\'eel
state is not an eigenstate of the system. However,
Anderson\cite{ander51} showed that the upper limit of the error
introduced by utilizing the classical Heisenberg model is
$1/(Z\,S)$, where $Z$ is the number of nearest neighbors. Since
the Eu-ions carry a relatively large spin of $S=7/2$, the error in
the ground state energy is smaller than 2.4\,\%. Thus, a classical
treatment considering continuously rotating spin vectors is well
justified.


\section{Exchange interactions in
E\lowercase{u}T\lowercase{e} at ambient pressure\label{sec:eute}}

In our Monte Carlo study of the exchange interactions, EuTe is
chosen as test material. This is because detailed experimental
data is available for the $H$-$T$ phase diagram\cite{olive72},
which allows a direct determination of the NN and NNN exchange
interaction $J_1$ and $J_2$ based on the antiferromagnetic
ordering temperature $T_N$ and the critical field at zero
temperature $H_C(T=0)$. In addition, the ferromagnetic ($J_1$) and
the antiferromagnetic exchange ($J_2$) are relatively balanced in
EuTe. Therefore, pronounced changes in magnetic ordering occur
when hydrostatic pressure is applied.

\subsection{Experimental Results}

The magnetic properties of EuTe were determined using DC and AC
SQUID magnetometry measurements of high quality 4\,$\mu$m thick
(111)-oriented epitaxial layers grown by molecular beam epitaxy on
BaF$_2$ substrates\cite{heiss01a,heiss01,heiss04}. As a result,
the magnetization and AC susceptibility was obtained as a function
of both temperature and external magnetic field up to 7\,T. The
external field direction was applied in the (111) growth plane,
which is also the easy plane of the magnetization.

\begin{figure}
  \includegraphics{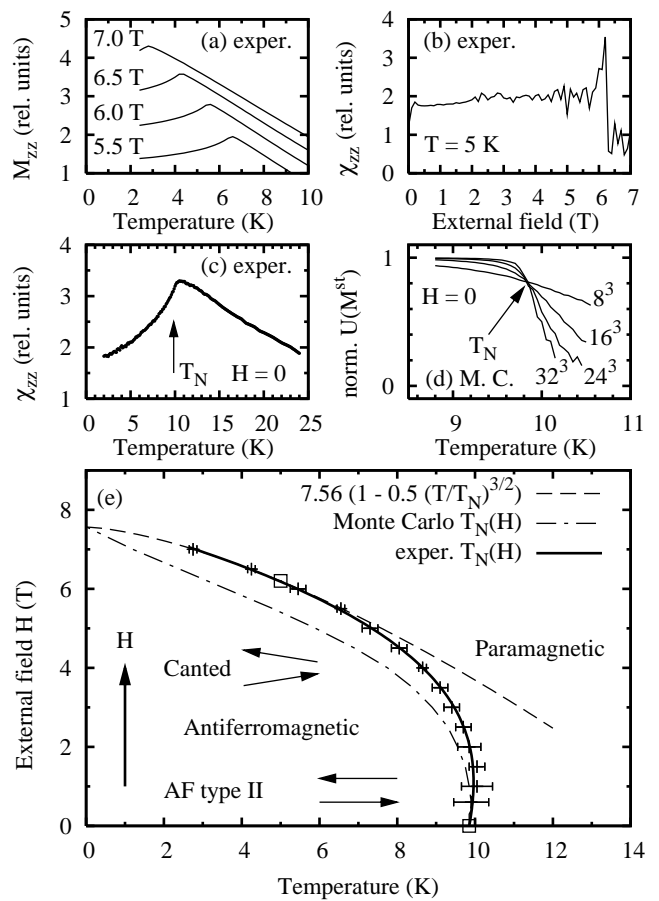}%
  \caption{Magnetic properties of bulk EuTe
at normal pressure derived from SQUID measurements and Monte Carlo
simulations. (a) Temperature-dependent magnetization $M(T)$
measured for various in plane external magnetic field values from
5.5 to 7 T. (b) AC susceptibility vs.\ external magnetic field at
5\,K indicating the critical field as discontinuity at $B \approx
6.2$\,T. (c) AC susceptibility $\chi$ measured as a function of
temperature at zero external field. The inclination point
indicated by the arrow yields a N\'eel temperature of
9.85$\pm$0.05\,K. (d) Determination of the N\'eel temperature from
Monte Carlo data using the fourth order cumulant of the staggered
magnetization $U(M^{st})$ for different system sizes. The curves
show a common intersection at the N\'eel temperature. (e) Phase
diagram of EuTe: Symbols with error bars correspond to maxima or
inflection points in $M(T)$ for high, respectively, small external
fields; squares to the discontinuity in $\chi(H)$ or the
inflection point in $\chi(T)$. The solid line indicates the
experimental $H_C(T)$ phase boundary and the dash-dotted line
represents the boundary obtained by the Monte Carlo calculations.
The $T^{3/2}$ extrapolation (dashed line) of the experimental data
towards $T=0$ yields a critical field of $H_C =
7.56$\,T.\label{fig:bulk_HTc}}
\end{figure}

Experimental magnetization curves $M(T)$ at different applied
external fields from 5.5 to 7 T are shown in
Fig.~\ref{fig:bulk_HTc}(a). The magnetization curves exhibit clear
peaks at the phase transition between the antiferromagnetic and
the paramagnetic phase. High magnetic fields were applied in order
to obtain information about the critical  field at zero
temperature $H_C(T=0)$. Field-dependent AC susceptibility curves
were also measured. As shown in Fig.~\ref{fig:bulk_HTc}(b), for
$T=5$\,K the susceptibility is essentially constant below the
critical field, corresponding to a linear increase in the
magnetization. This arises from the continuous reduction of the
relative angle between the spins in the adjacent (111) lattice
planes from 180° in the AFM II state at zero field to almost zero
at the critical field $H_C$, as illustrated schematically by the
arrows in Fig.~\ref{fig:bulk_HTc}(e). Thereby the spin orientation
changes from initially perpendicular to the external field to
finally parallel to the external field direction when $H$ reaches
the critical field. The critical field is thus given by the
discontinuous drop in the susceptibility at this point (see
Fig.~\ref{fig:bulk_HTc}(b)).

Figure~\ref{fig:bulk_HTc}(c) displays the susceptibility as a
function of temperature at zero external field, showing a broad
peak with a maximum slightly above 10\,K. However, the phase
transition from antiferromagnetic to paramagnetic corresponds to
the peak in the specific heat, which according to Fisher's
rule\cite{fishe62} coincides with the maximum slope, i.e., the
inclination point in the AC susceptibility. As indicated by the
arrow in Fig.~\ref{fig:bulk_HTc}(c), the phase transition thus
occurs at $T_N = 9.85 \pm 0.05$\,K.

Figure~\ref{fig:bulk_HTc}(d) shows the phase diagram of EuTe
compiled from the experimental data (symbols and solid line),
where the thick solid line indicates the experimentally determined
$H_C(T)$ phase boundary. A $T^{3/2}$ curve, which according to spin wave
theory is the low-temperature behavior of the critical
field, was fitted to the experimental critical points between 2\,K
and 6\,K, yielding
\begin{equation}\label{eqn:t32law}
 H_C(T) = H_C(0)\left[1-\lambda(T/T_N)^{3/2}\right]\quad.
\end{equation}
Extrapolation of the the measured $H_C(T)$ to $T=0$ thus yields
the critical  field at zero temperature of
$H_C(0)=7.56\pm0.02$\,K. The dashed line in
Fig.~\ref{fig:bulk_HTc}(e) represents Eqn.~(\ref{eqn:t32law}) with
the coefficient $\lambda = 0.50\pm0.01$. The experimental data as
well as $T_N$, $H_C(0)$ and $\lambda$ are in excellent agreement
with previous results of Oliveira et al.\cite{olive72}.

\subsection{Exchange constants and phase diagram from MC calculations}

\begin{table}

 \begin{ruledtabular}
 \begin{tabular}{lccrr}
Reference              & Exper. results / Analysis     & $J_1$ (K) & $J_2$ (K)\\
\hline
Oliveira\cite{olive72} & $H_C$, $\theta$ / MFA         & 0.100        & -0.215\\
Zinn\cite{zinn76}      & $H_C$, $\theta$ / MFA         & 0.060        & -0.200\\
Wachter\cite{wacht79}  & $H_C$, $T_N$    / MFA         & 0.043        & -0.150\\
K\"obler\cite{koebl93} & $H_C$, $T_N$    / MFA         & 0.060        & -0.160\\
Kune\v{s}\cite{kunes05}& ab initio LDA+$U$ ($U=6$\,eV) & 0.110        & -0.320\\
\hline
Our study              & $H_C$, $T_N$ / MFA            & 0.035        & -0.156\\
                       & $H_C$, $T_N$ / MC             & 0.192        & -0.313
\end{tabular}
\end{ruledtabular}
\caption{Comparison of the exchange constants $J_1/k_B$ and
$J_2/k_B$ of EuTe determined by the analysis of experimental data
for the N\'eel temperature $T_N$, the critical field $H_C$ or the
paramagnetic Curie temperature $\theta_C$ using the mean field
approximation (MFA) or the Monte Carlo (MC) method (present work).
Also listed are the exchange constants derived by Kune\v{s} et al.
from ab initio calculations.\label{tab:exchange_constants}}
\end{table}

In most previous studies\cite{olive72,zinn76,wacht79,koebl93}, the
mean field analysis was used to determine the exchange integrals
in EuTe, because it provides simple analytic expressions for the
critical field at zero temperature $H_C(0)$, the N\'eel
temperature $T_N$ as well as the paramagnetic Curie temperature
$\theta_C$ as a function of $J_1$ and $J_2$. For
type~II-antiferromagnetic ordering in an \textit{fcc} lattice with
NN and NNN exchange interactions, the mean field approximation
(MFA) yields
\begin{eqnarray}
H_C^{\text{MFA}}(0) & = & -4\,S\,(6\,J_1+6\,J_2)/(g\,\mu_B)\quad,
                \label{eqn:mfa_crit_fld}\\
   T_N^{\text{MFA}} & = & \frac{2}{3}\,S\,(S+1)\,(-6\,J_2)/k_B\quad,
                \mbox{and}\label{eqn:mfa_neel_temp}\\
\theta_C^{\text{MFA}} &=& \frac{2}{3}\,S\,(S+1)\,(12\,J_1+6\,J_2)/k_B\quad,
                \label{eqn:mfa_para_curie_temp}
\end{eqnarray}
where $g = 2$ and $S = 7/2$ for the magnetic moment of the
Eu\textsuperscript{2+} ions. Inserting our experimental values for
$H_C(0)$ and $T_N$ into Eqns.~(\ref{eqn:mfa_crit_fld})
and~(\ref{eqn:mfa_neel_temp}) and solving for $J_1$ and $J_2$
yields $J_1^{\text{MFA}}/k_B=0.035$\,K and
$J_2^{\text{MFA}}/k_B=-0.156$\,K. As shown in
Tab.~\ref{tab:exchange_constants}, the values are consistent with
previous mean field studies, especially those, which applied the
same analysis of the experimental $H_C(0)$ and $T_N$ based on MFA
Eqns.~(\ref{eqn:mfa_crit_fld}) and~(\ref{eqn:mfa_neel_temp}).

In the Monte Carlo calculations, the transition temperature
$T_N(0)$ ($T_N(H\neq0)$) is deduced from the temperature
dependence of the  4\textsuperscript{th} order cumulants of the
staggered (transverse) magnetization $U^{st}_{(\bot)}$. As
described in Sec.~\ref{sec:computational}, these cumulants show
universal values at the critical temperature $T_N$ independent of
cluster size. This is demonstrated in Fig.~\ref{fig:bulk_HTc}(d)
for the case of $H=0$. Using the exchange constants $J_1^{MFA}$
and $J_2^{MFA}$ derived from the mean field analysis, the Monte
Carlo calculations yield a N\'eel temperature of 5.45\,K at zero
external field, which is in strong disagreement with the
experimental value of 9.85\,K. This clearly demonstrates that the
neglection of spin fluctuations in MFA leads to a vast
underestimation of the exchange constants, an effect that has been
already noted in previous theoretical studies
\cite{nowot89,scott02}. In the MC calculations, moreover, the
critical N\'eel temperature $T_N$ is found to depended
significantly not only on the antiferromagnetic exchange constant
$J_2$ but also on the ferromagnetic NN exchange constant $J_1$, in
contrast to the MFA approximation where $T_N$ depends only on the
antiferromagnetic exchange ---see Eqn.~(\ref{eqn:mfa_neel_temp}).
This is due to the fact that in the mean field approximation, for
type~II antiferromagnetic ordering the $J_1$ exchange between the
6 NN Eu\textsuperscript{2+} ions within the ferromagnetic (111)
planes exactly cancels with the $J_1$ exchange to the 6 NN
Eu\textsuperscript{2+} ions  within the antiferromagnetically
coupled neighboring (111) lattice planes. This does not apply for
the MC calculations because of the non perfect antiferromagnetic
spin alignment at nonzero temperatures that results from spin
fluctuations.

\begin{figure}
  \includegraphics{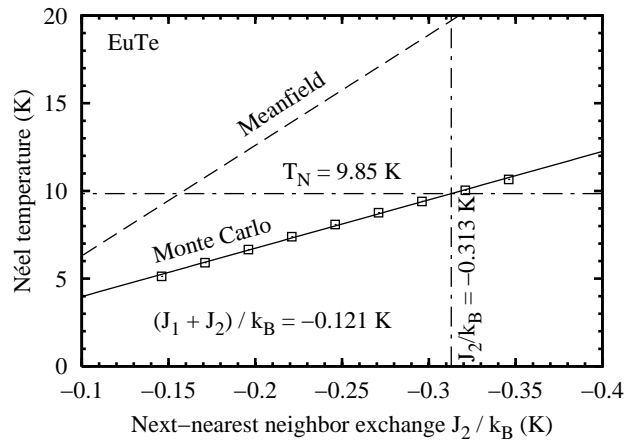}%
  \caption{N\'eel temperature of EuTe as a function of the
nearest neighbor exchange constant $J_1$ under the constraint
$(J_1 + J_2)/k_B = -0.121$\,K, which is a consequence of
evaluating the experimental critical field at zero temperature
$H_C(T=0)$---see in the text for details. We obtain a linear
function (squares and solid line) which is shown in comparison
with the corresponding mean field relation (dashed line). At
$J_2=-0.313$\,K ($J_1 = 0.192$\,K) the Monte Carlo curve reaches
the experimental N\'eel point of
9.85\,K.\label{fig:eute_bulk_TnJi_lin}}
\end{figure}

In order to determine the exchange constants from the Monte Carlo
calculations, we have systematically calculated the N\'eel
temperature as a function of both exchange constants $J_1$ and
$J_2$. As a boundary condition, we take advantage of the fact that
in the limit of $T=0$, the critical field $H_C$ of the Monte Carlo
calculations converges exactly to the mean field value of
Eqn.~(\ref{eqn:mfa_crit_fld}). This is because spin fluctuations
are absent at $T=0$, and thus the MFA represents the exact ground
state of the system. Therefore, the experimental value of
$H_C(T=0)$ and Eqn.~(\ref{eqn:mfa_crit_fld}) determine the sum of
the exchange constants as
\begin{equation}\label{eqn:sum_J_1_J_2}
 (J_1+J_2)/k_B = -0.121\,\mbox{K}.
\end{equation}
in EuTe. This eliminates one independent variable, i.e., with
this condition, only $J_2$ must be varied for the calculation of
$T_N$. The resulting dependence is plotted in
Fig.~\ref{fig:eute_bulk_TnJi_lin} (squares and solid line).
Evidently, $T_N$ varies  almost perfectly linearly and can be
represented by the relation
\begin{equation}\label{eqn:mc_neel_temp_constraint}
 \left.T_N^{\text{MC}}\right|_{(J_1+J_2)/k_B = -0.121\,\text{K}} =
1.22-27.57\,J_2/k_B
\end{equation}
within the range $-0.35 \leq J_2/k_B \leq -0.15$. Solving
Eqn.~(\ref{eqn:mc_neel_temp_constraint}) for $J_2$ and inserting
the experimental N\'eel point of bulk EuTe $T_N = 9.85$\,K
(horizontal dash-dotted line in Fig.~\ref{fig:eute_bulk_TnJi_lin})
yields $J_2^{\text{MC}}/k_B = -0.313$\,K and hence
$J_1^{\text{MC}}/k_B = 0.192$\,K from Eqn.~(\ref{eqn:sum_J_1_J_2})
as the intrinsic exchange constants of bulk EuTe. It is noted,
that calculating $T_N$ without the constraint of
Eqn.~(\ref{eqn:sum_J_1_J_2}) yields a function $T_N(J_1,J_2)$ that
depends nonlinearly on $J_1$ and $J_2$, in contrast to the MFA
Eqn.~(\ref{eqn:mfa_neel_temp}), which predicts only a linear
dependence on $J_2$---see Sec.~\ref{sec:distance_eute} for further
details. From the MC calculations, the N\'eel temperature as a
function of $J_1$ and $J_2$ is found to be well described by
\begin{eqnarray}\label{eqn:mc_neel_temp}
 T_N^{\text{MC}} & \approx & (- 15.3\,J_1 - 40.8\,J_2)/k_B \nonumber\\
          &    =    & \frac{2}{3}\,S\,(S+1)\,(- 1.46\,J_1 - 3.89\,J_2)/k_B
\end{eqnarray}
in a linear approximation in the vicinity of the  intrinsic EuTe
exchange constants, demonstrating that the N\'eel temperature
indeed depends strongly on both exchange constants.

As demonstrated by Tab.~\ref{tab:exchange_constants}, which
compares our derived set of exchange constants with previously
published ones, our values are nearly twice as large as those
derived from mean field analysis. Thus, by neglection of spin
fluctuations the exchange parameters are vastly underestimated.
Remarkably, the exchange constants derived from our Monte Carlo
calculations are in good agreement with recent ab initio
calculations of Kune\v{s} et al.\cite{kunes05} using the
local-density-approximation method including  strong Coulomb
repulsion within the 4$f$ shells (LDA+$U$). In particular, our NNN
exchange constant $J_2$, which in Ref.~\onlinecite{kunes05}
depends very weakly on the Coulomb parameter $U$ matches the ab
initio result very well.

With the new exchange parameters, we can now calculate the whole
magnetic phase diagram of EuTe using the 4\textsuperscript{th}
order cumulants of the staggered (transverse) magnetization for
various external magnetic fields and cluster sizes.  The resulting
phase boundary $T_N(H)$  is depicted as dash-dotted line in
Fig.~\ref{fig:bulk_HTc}(e). As expected, the calculated $H_C(T)$
approaches the experimental value of 7.56\,T in the limit of $T
\rightarrow 0$ and the calculations nearly follow the experimental
$H_C(T)$ boundary. The fact, that at $T>0$ the calculated $H_C(T)$
values are slightly lower than the measured ones and that at low
temperatures, the calculated critical field varies linearly with
temperature instead of obeying a $T^{3/2}$ behavior is a well
known consequence of applying a classical ($S=\infty$) instead of
the quantum mechanical $S=7/2$ model in our calculations. As
already noted in Sec.~\ref{sec:computational}, the error in the
ground state energy introduced by this simplification is of the
order of less than 2.4\,\% for our type of system.

\section{Distance dependence of exchange interactions in
E\lowercase{u}T\lowercase{e}\label{sec:distance_eute}}

\begin{figure}
  \includegraphics{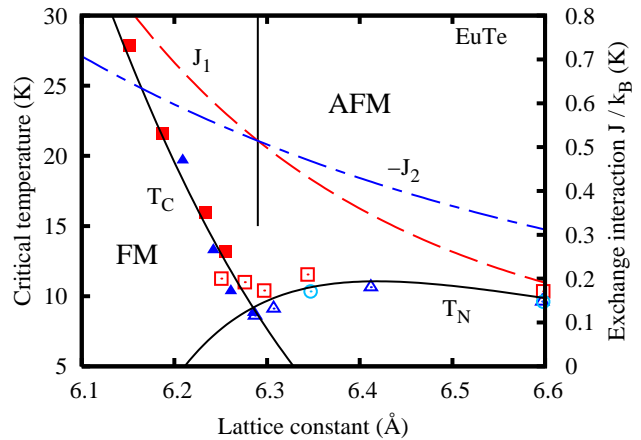}%
  \caption{(color online) Critical temperatures (left scale) and exchange
interactions $J_1$ and $J_2$ (right scale) in EuTe plotted as a
function of the lattice constant $a$. Open and closed symbols
(circles\cite{klein77}, squares\cite{gonch97}, triangles\cite{ishiz97}),
correspond to N\'eel ($T_N$), respectively
Curie points ($T_C$) determined from hydrostatic pressure
experiments. Solid lines: $T_N(a)$ and $T_C(a)$ derived from Monte
Carlo calculations using a $J_1(a)$ and $J_2(a)$ dependence given
by the Gr\"uneisen law (\ref{eqn:grueneisen}) with best fitting
power law exponents of $n_1 = 20.6\pm0.4$ and $n_2 = 10.4\pm0.5$,
respectively. The corresponding variation of $J_1(a)$ and $J_2(a)$
as functions of the lattice parameter is shown as the dashed and
dash-dotted line, respectively. \label{fig:xcfit_EuTe}}
\end{figure}

Under hydrostatic pressure, the EuTe lattice constant can be
compressed from its normal bulk value of $a_0$ = 6.589\,\AA\ to
about 6.15\,\AA\ at a pressure reaching 17\,GPa \cite{gonch97}. This
corresponds to a 7\,\% reduction of the lattice constant and of
the interatomic distances of the Eu\textsuperscript{2+} ions in
the crystal , where in the $fcc$ lattice of EuTe the NN
Eu\textsuperscript{2+} distance is $r_1 = a/\sqrt{2}$ and the NNN
distance $r_2 = a$. The resulting changes in the magnetic ordering
temperatures obtained by hydrostatic pressure experiments
\cite{klein77,ishiz97,gonch97} are compiled in
Fig.~\ref{fig:xcfit_EuTe}, where the open symbols represent the
measured $T_N$ and the full symbols the $T_C$ values plotted as a
function of lattice constant. Since at ambient pressure, the
antiferromagnetic exchange $J_2$ of EuTe is larger than the
ferromagnetic exchange $J_1$, a type~II antiferromagnetic
ordering\cite{seehr88} occurs below the N\'eel point of $T_N$ =
9.85 K. With increasing pressure, i.e., decreasing lattice
constant, the N\'eel temperature remains practically constant at
 $T_N \approx 10$\,K, but at $\approx 9$\,GPa or 5\,\%
compressive strain EuTe becomes ferromagnetic\cite{ishiz97,
gonch97} with rapidly increasing Curie $T_C$ that rises up to
28\,K at 17\,GPa\cite{gonch97}. The observed phase transition from
antiferromagnetism to ferromagnetism at $a = 6.29$\,\AA\ implies
that at smaller atom distances the NN $J_1$ becomes the dominating
exchange mechanism.

The influence of the inter-atomic distances $r_i$ on exchange
constants has been a subject of many theoretical
studies\cite{kasuy70,zinn76,lee84,khoms97,gonch00,radom01}. However,
 indirect and superexchange mechanisms involve complex
integrals such that up to now no general analytic expressions for
their distance dependence have been  derived theoretically. An
empirical power law dependence, referred to as the magnetic
Gr\"uneisen law, has been proposed by Bloch\cite{bloch66}, i.e.,
\begin{equation}\label{eqn:grueneisen}
 J(r) = J_0\left(\frac{r}{r_0}\right)^{-n}\quad,
\end{equation}
where $J_0=J(r_0)$ and $r_0$ are the exchange interaction and
interatomic distance at normal pressure, and $n$ is the scaling
exponent. As shown in Ref.~\onlinecite{bloch66}, this dependence
well describes the observations for many magnetic semiconductors
or insulators such as the Mn and Gd chalcogenides or iron oxides,
for which the power law exponent $n$ shows a universal value of
around 10 for the magnetic superexchange\cite{bloch66}. This also
yields the empirical 10/3 law for the volume dependence of
superexchange\cite{bloch66} of $J(V) = J_0(V/V_0)^{-10/3}$.

\begin{figure*}
\includegraphics{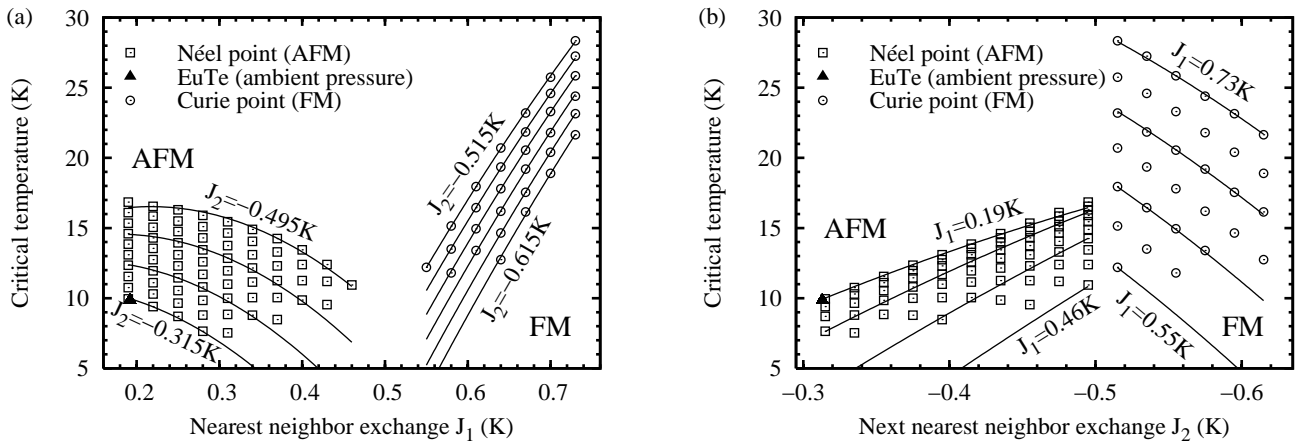}
\caption{\label{fig:eute_bulk_TnJi}Monte Carlo calculated critical
temperatures of a bulk \textit{fcc} Heisenberg system as functions
of the strength of (a) the nearest neighbor exchange interaction $J_1$ and
(b) the next-nearest neighbor exchange interaction $J_2$, showing
antiferromagnetic to paramagnetic transitions (squares) when
$J_1<|J_2|$ and ferromagnetic to paramagnetic transitions
(circles) when $J_1>|J_2|$. Solid lines represent polynomial fit
functions, which are quadratic in $J_1$ and $J_2$.}
\end{figure*}

To test if the Gr\"uneisen dependence of
Eqn.~(\ref{eqn:grueneisen}) adequately describes the atomic
distance dependence of the exchange integrals in EuTe, we have
performed a series of Monte Carlo calculations of the N\'eel and
Curie temperature as a function of the exchange integrals in order
to fit the experimental $T_N(a)$ and $T_C(a)$ data of
Fig.~\ref{fig:xcfit_EuTe} using the power law exponents $n_1$ and
$n_2$ for the NN and NNN exchange interactions $J_1$ and $J_2$ as
free parameters. In these calculations, the ferromagnetic NN
exchange $J_1>0$ and the antiferromagnetic NNN  exchange $J_2<0$
were varied independently in the range of $0.190 \leq J_1 \leq
0.73$ and $0.315 \leq -J_2 \leq 0.615$ and the corresponding
critical temperatures were derived as described in detail in
Secs.~\ref{sec:computational} and~\ref{sec:eute}. For all $|J_1 |
< | J_2 |$, the MC calculations yield AFM~II ordering, whereas FM
ordering results for all $| J_1 |
> | J_2 |$. Figures~\ref{fig:eute_bulk_TnJi}(a) and (b) show the
calculated $T_N$ and $T_C$ values (open symbols) as a function of
$J_1$ and $J_2$, respectively. Since we are interested in the
hydrostatic pressure effect, the smallest values of $J_1$ and
$J_2$ were chosen close to the exchange parameters of bulk EuTe at
ambient pressure (filled symbols) and the maximum values
correspond to hydrostatic pressures up to about 17\,GPa.

As shown by Fig.~\ref{fig:eute_bulk_TnJi}, the Monte Carlo
calculations yield a strongly nonlinear dependence of $T_N$ on
$J_1$. Hence, the calculated data points were approximated by
second degree polynomials for $T_N(J_1,J_2)$ and $T_C(J_1,J_2)$ of
the general form
\begin{equation}\label{eqn:poly_biquad}
T(J_1, J_2)=A+BJ_1+CJ_2+DJ_1^2+EJ_1J_2+FJ_2^2\quad.
\end{equation}
These approximations fit the Monte Carlo data with better than
$\pm0.1$\,K accuracy and are represented as solid lines in
Figs.~\ref{fig:eute_bulk_TnJi}(a) and (b).

Using the normal pressure exchange constants $J_{1,0}=0.192$\,K
and $J_{2,0}=-0.313$\,K determined in Sec.~\ref{sec:eute}, the two
branches of experimental data sets for the antiferromagnetic
$T_N(a)$ and ferromagnetic $T_C(a)$ of Fig.~\ref{fig:xcfit_EuTe}
were fitted using Eqns.~(\ref{eqn:grueneisen}) and
(\ref{eqn:poly_biquad}) with common exponents $n_1$ and $n_2$ and
$r_1 = a/\sqrt(2)$ and $r_2 = a$. All experimental data points are
weighted equally in the least square fit routine, which was
performed on a logarithmic scale since equal weights may cause one
branch to dominate if there is a difference in the magnitude of
the dependent variable.

From this modeling, a Gr\"uneisen exponent of $n_1 = 20.6\pm0.4$
for the NN exchange $J_1$ and of $n_2 = 10.4\pm0.5$ for the NNN
exchange $J_2$ were obtained. As demonstrated by the solid lines
in Fig.~\ref{fig:xcfit_EuTe}, with these parameters the whole body
of experimental findings, including the approximately constant
N\'eel temperature $T_N$ at small hydrostatic strain, the
transition from antiferromagnetic to ferromagnetic ordering at $a
= 6.29$\,\AA\ and the steep superlinear increase of the Curie
temperature at high  pressures, i.e., small lattice constants,
are exactly reproduced. Moreover, the obtained power law exponent
$n_2$ for $J_2(r_2)$ is in excellent agreement with
Bloch's\cite{bloch66} 10/3 law for the volume dependence of
superexchange.

The such obtained dependence of the NN  and NNN exchange integrals
$J_1(a)$ and $J_2(a)$ as a function of lattice parameter $a$ is
presented in Fig.~\ref{fig:xcfit_EuTe} as dashed and dash-dotted
line, respectively. Evidently, both exchange constants strongly
increase with decreasing lattice constant. However, $J_1(a)$
increases much more rapidly than $J_2(a)$ due to the two times
larger power law exponent. Therefore, the two curves intersect at
$a=6.29$\,\AA, where $J_1/k_B = -J_2/k_B = 0.51$\,K, and at
smaller $a$, the ferromagnetic $J_1$ becomes the dominating
exchange mechanism. For \textit{fcc} lattices with competing
ferromagnetic NN exchange and antiferromagnetic NNN exchange
interactions, this is exactly the condition for the material to
become ferromagnetic\cite{seehr88}.

Our results are in severe contrast to the previous mean field
analysis of Goncharenko et al.\cite{gonch97}, who concluded from
the negligible variation of $T_N$ in the antiferromagnetic state
with changing lattice constant that the NNN exchange $J_2$ in
Eu$X$ should \emph{not} depend on the lattice parameter.
Consequently, the whole variation of $T_C(a)$ was attributed
solely to changes in $J_1(a)$ using the mean field expression for
the Curie temperature of
\begin{equation}\label{eqn:mfa_curie_temp}
 T_C^{\text{MFA}} = \frac{2}{3}\,S\,(S+1)\,(12\,J_1+6\,J_2)/k_B\quad.
\end{equation}
with constant $J_2$ for data analysis. On the contrary, our
calculation show that the broad plateau of $T_N(a)$ for lattice
constants around $a=6.42$\,\AA\ just results from the fact that in
the antiferromagnetic phase the ferromagnetic exchange drops
faster than the antiferromagnetic exchange as the lattice constant
increases.

\section{Exchange interactions in E\lowercase{u}O and
E\lowercase{u}S\label{sec:euo_eus}}

As shown in the previous sections, the exchange constants
obtained by Monte Carlo calculations strongly differ from
previously published values. Therefore, to evaluate the distance dependence
of the exchange constants of EuO and EuS, first the bulk values
under ambient pressure have to be reexamined by the
Monte Carlo method.

\subsection{Exchange constants at ambient pressure\label{sec:bulk_euos}}

EuO and EuS are low temperature ferromagnets with Curie
temperatures $T_C$ of 69.15\,K\cite{passe76} and
16.6\,K\cite{passe76,bohn81}, respectively. The magnetic
properties are determined mainly by the dominant ferromagnetic NN
exchange interaction $J_1$ in both materials. Compared to the case
of antiferromagnetic EuTe, where  only the N\'eel temperature and
the critical field at $T = 0$\,K are needed to deduce the exchange
constants, the determination of $J_1$ and $J_2$ for ferromagnetic
EuO and EuS is much more involved. As a result, there exists a
substantial variation in the reported exchange constants for bulk
EuO and EuS deduced from different experimental techniques like
inelastic neutron scattering\cite{passe76,bohn81,mook81}, specific
heat\cite{passe66,hende70,schwo71,dietr75}, nuclear magnetic
resonance\cite{chara64,boyd66,comme05} (NMR) and spin wave
resonance measurements\cite{schwo74} (see, e.g., Passell et
al.\cite{passe76} for a review). Especially the values for the NNN
exchange constant $J_2$ in EuO and EuS differ by up to a factor of
two in literature and whether $J_2$ is ferromagnetic or
antiferromagnetic in EuO is still a matter of debate. The most
recent results based on inelastic neutron scattering studies on
single crystals of EuO\cite{mook81} and EuS\cite{bohn81} yielded
$(J_{1,0}^{\text{EuO}}/k_B, J_{2,0}^{\text{EuO}}/k_B) =
 (0.625\,\mbox{K},          0.125\,\mbox{K})$ and
$(J_{1,0}^{\text{EuS}}/k_B, J_{2,0}^{\text{EuS}}/k_B) =
 (0.221\,\mbox{K},         -0.100\,\mbox{K})$, respectively,
consistent with Passell et al.'s\cite{passe76} analysis
on powdered samples. Notably, a ferromagnetic NNN exchange interaction
was obtained for EuO.

In all studies, the sum of $J = J_{1}+J_{2}$ has been more
reliably determined than the individual NN and NNN exchange
interactions, and this sum is quite consistent among the various
studies. For EuO single crystals, $J^{\text{EuO}}/k_B = 0.755$\,K
was obtained by Comment et al.\cite{comme05} from NMR
measurements, in agreement with neutron scattering studies by Mook
at al.\cite{mook81}, and this value also agrees with the results
obtained by neutron scattering\cite{passe76} and specific heat
measurements \cite{dietr75} on powdered samples. A very good
agreement for $J_1 + J_2$ also exist among respective studies for
EuS\cite{chara64,bohn81,passe76,dietr75}, from which we calculate
$J/k_B^{\text{EuS}} = 0.121\pm0.003$\,K as  mean value.

\begin{figure}
  \includegraphics{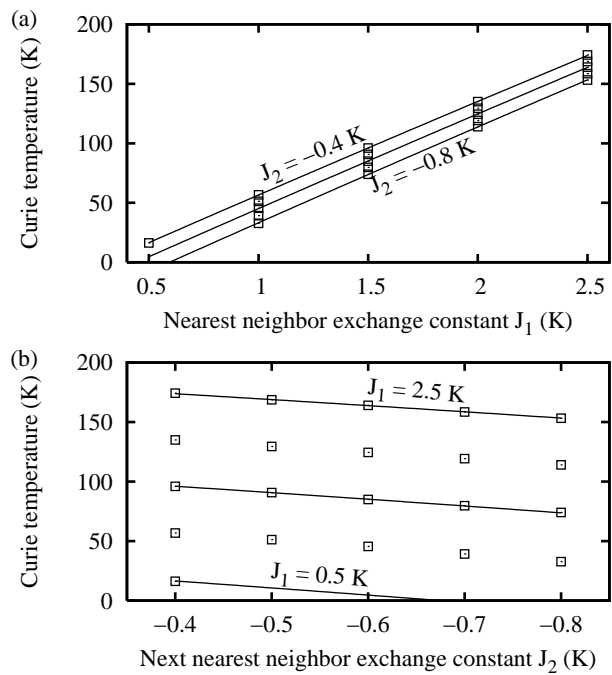}%
  \caption{Monte Carlo calculated critical temperatures
of a bulk \textit{fcc} Heisenberg system as functions of the strength of (a)
the nearest and (b) next-nearest neighbor exchange interactions ($J_1$ and
$J_2$, respectively), showing ferromagnetic to paramagnetic
transitions if $J_1>|J_2|$. The parameter range covers EuO and EuS
under hydrostatic pressures between 0 and 20\,GPa. Solid lines
represent fit functions, which are linear in $J_1$ and $J_2$---see
Eqn.~(\ref{eqn:mc_curie_temp}).
\label{fig:eus_bulk_TcJi}}
\end{figure}

To determine the exchange constants by the Monte Carlo method, we
again performed a series of calculations for the model Heisenberg
Hamiltonian of Eqn.~(\ref{eqn:hamiltonian}) with the NN and NNN
exchange interactions $J_1$ and $J_2$ varied independently over a
wide range of $0.5 \leq J_1 \leq 2.5$ and $0.4 \leq -J_2 \leq
0.8$. We find ferromagnetic ordering for all combinations of $J_1
> -J_2$ and determined the corresponding critical ordering temperature
as a function of $J_1$ and $J_2$ as  presented in
Fig.~\ref{fig:eus_bulk_TcJi} (a) and (b), respectively. Evidently,
$T_C$ increases linearly with increasing NN exchange constant
$J_1$, but \emph{decrease}  when $J_2$ increases. As shown by the
solid lines in Fig.~\ref{fig:eus_bulk_TcJi}, the general
dependence of $T_C$ on the exchange constants can be well
described by the relation
\begin{eqnarray}\label{eqn:mc_curie_temp}
 T_C^{\text{MC}} & = & 79.0\,J_1/k_B + 55.9\,J_2/k_B \\
                 & = &
0.627\,\frac{2}{3}\,S\,(S+1)\,12\,(J_1+0.708\,J_2)/k_B\quad, \nonumber
\end{eqnarray}
which differs considerably compared to the mean field expression
of Eqn.~(\ref{eqn:mfa_curie_temp}). Yet, the critical coupling
$J_1/(k_B T_C)$ closely resembles the theoretical predictions from
high-temperature series expansion\cite{ritch72}. Inserting the EuO
and EuS exchange values of Mook\cite{mook81} and Bohn et
al.\cite{bohn81} in our Monte Carlo relation of
Eqn.~(\ref{eqn:mc_curie_temp}) yields Curie temperatures of only
56\,K and 12\,K, respectively, which is much lower than the
measured experimental values. This shows that, like for EuTe, the
exchange constants have been considerably underestimated in both
materials.

\begin{figure*}
  \includegraphics{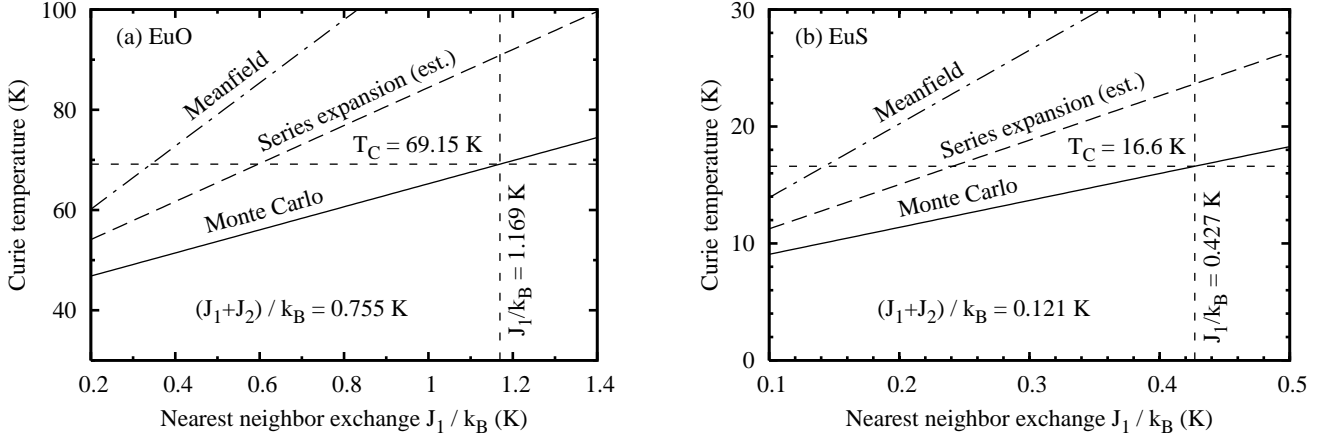}%
  \caption{Determination of the intrinsic nearest neighbor exchange constant
$J_{1,0}$ in (a) EuO and (b) EuS from Monte Carlo results (solid
lines) under the constraint $J = J_1 + J_2 = \mbox{const.}$.
Experimental values for $J$ are taken from
Refs.~\onlinecite{comme05} (EuO) and~\onlinecite{bohn81} (EuS).
The Curie temperature as functions of $J_1$ as derived from mean
field approximation (dash-dotted line) and an approximate series
expansion results\cite{passe76} (dashed line) are shown in
comparison. The dashed horizontal lines indicate the measured bulk
$T_C$ values of 69.15 and 16.6\,K for EuO and EuS, respectively.
\label{fig:euo_eus_bulk_TcJi_lin}}
\end{figure*}

Using the  experimental values for $J = J_{1}+J_{2}$ quoted above,
the critical Curie temperature $T_C$ can be calculated as a
function of the exchange constant $J_{1}$ using
Eqn.~(\ref{eqn:mc_curie_temp}) and $J_{2} = J - J_{1}$. The
results are plotted as solid lines in
Fig.~\ref{fig:euo_eus_bulk_TcJi_lin}(a) and~(b) for EuO
($J^{\text{EuO}} = 0.755$\,K) and EuS ($J^{\text{EuS}} =
0.121$\,K), respectively. From the intersection of these lines
with the respective experimental $T_C$ values of
$69.15$\,K\cite{passe76} and $16.6$\,K\cite{passe76,bohn81}
(horizontal dashed lines in Fig.~\ref{fig:euo_eus_bulk_TcJi_lin}),
the bulk exchange constants of $J_{1}^{\text{EuO}} = 1.169$\,K and
$J_{2}^{\text{EuO}} = -0.414$\,K are obtained for EuO and of
$J_{1}^{\text{EuS}} = 0.427$\,K and $J_{2}^{\text{EuS}} =
-0.306$\,K for EuS. In Fig.~\ref{fig:euo_eus_bulk_TcJi_lin}, also
plotted are the $T_C$ values expected from the mean field
approximation (dash-dotted line, Eqn.~(\ref{eqn:mfa_curie_temp}))
as well as from a series expansion estimate proposed by Passell et
al.\cite{passe76} (dashed line)  given by
\begin{equation}\label{eqn:se_curie_temp}
 T_C^{\text{SE, est.}} =
0.790\,\frac{2}{3}\,S\,(S+1)\,12\,(J_1+0.619\,J_2)/k_B\quad.
\end{equation}
Evidently, in both cases, much higher critical temperatures are
predicted for a given set of exchange constants, i.e., from the
observed transition temperatures, the exchange integrals are
strongly underestimated. Moreover, a ferromagnetic NNN exchange
would be suggested for EuO. Our results  exclude such a
ferromagnetic NNN exchange, i.e., a negative $J_{2}$ is obtained
for EuO even if $J^{\text{EuO}}$ is underestimated by as much as
15\,\%. Thus, as already found for the EuTe case, our Monte Carlo
calculations greatly revise the bulk exchange constants.

\subsection{Distance dependence of EuO and EuS exchange
constants\label{sec:distance_euos}}

\begin{figure}
  \includegraphics{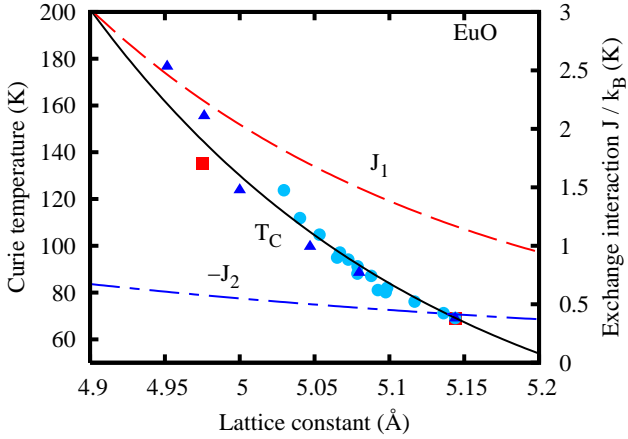}%
  \caption{(color online) Critical temperatures and exchange
interactions in EuO as functions of the lattice parameter;
filled symbols (circles\cite{mcwha66},
triangles\cite{moser88}, squares\cite{abdel90}) correspond to experimental Curie
points; the solid line represents the least square fit of Monte Carlo critical
temperatures $T_C(J_1,J_2)$ based on the magnetic Gr\"uneisen law $J_i(r_i) \sim
r_i^{-n_i}$. The corresponding dependence of the nearest ($J_1$) and
next-nearest neighbor
exchange interaction ($J_2$) on the lattice parameter is plotted as the dashed
and dash-dotted line, respectively. \label{fig:xcfit_EuO}}
\end{figure}

\begin{figure}
  \includegraphics{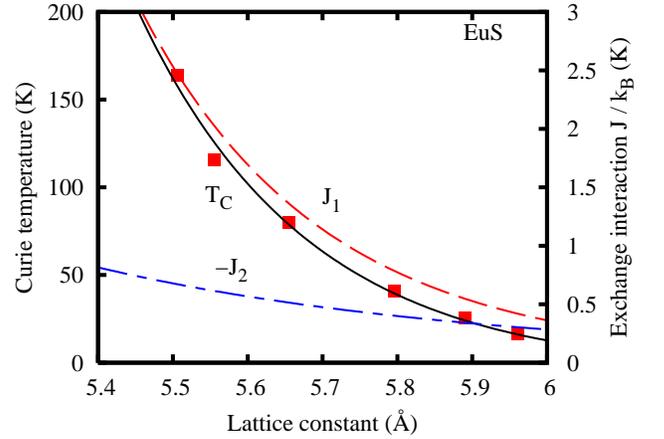}%
  \caption{(color online) Critical temperatures and exchange
interactions in EuS as functions of the lattice parameter;
squares\cite{gonch98} correspond to experimental Curie points;
the solid line represents the least square fit of Monte Carlo critical
temperatures $T_C(J_1,J_2)$ based on the magnetic Gr\"uneisen law $J_i(r_i) \sim
r_i^{-n_i}$. The
corresponding dependence of the nearest ($J_1$) and next-nearest neighbor
exchange
interaction ($J_2$) on the lattice parameter is plotted as the dashed and
dash-dotted line, respectively. \label{fig:xcfit_EuS}}
\end{figure}

For EuO and EuS, the Curie temperature $T_C$ strongly increases
with increasing hydrostatic pressure, i.e., decreasing lattice
constant. This is illustrated by Figs.~\ref{fig:xcfit_EuO}
and~\ref{fig:xcfit_EuS}, where the experimentally determined $T_C$
values of EuO\cite{mcwha66,moser88,abdel90} and EuS\cite{gonch98}
are plotted as a function of the lattice constant. At hydrostatic
pressures around 20 GPa, corresponding to a 6--8\,\% reduction of
the lattice constant, $T_C$ is as high as 200\,K\cite{abdel90} and
180\,K\cite{gonch98}, respectively. For EuO, the experimental data
$T_C(a)$ of Fig.~\ref{fig:xcfit_EuO} is compiled from three
independent investigations\cite{abdel90,moser88,mcwha66}, with
McWhan et al.'s data combined with the pressure-volume relation
taken from Ref.~\onlinecite{heath95}. For EuS, the data is taken
from Ref.~\onlinecite{gonch98}. For EuO, experiments at even
higher hydrostatic pressures up to 31\,GPa\cite{abdel90} have
revealed that the ferromagnetic ground state becomes unstable at
around 23\,GPa and that the Curie temperature drops sharply
afterwards instead of further increasing. In
Ref.~\onlinecite{abdel90} this behavior was attributed to
\textit{sf} hybridization competing with \textit{sf} exchange in
this pressure range. Therefore, we restrict our analysis to the
0--20\,GPa range, i.e., lattice constants above 4.9\,\AA, where
such effects seem not to be of importance.

To determine the interatomic distance dependence of the exchange
constants, we proceed in the same manner as described in
Sec.~\ref{sec:distance_eute} by fitting the calculated
$T_C^{\text{MC}}(J_1, J_2)$ dependence of
Eqn.~(\ref{eqn:mc_curie_temp}) obtained by the Monte Carlo
calculations to the data of the hydrostatic pressure experiments,
applying the magnetic Gr\"uneisen law
(Eqn.~(\ref{eqn:grueneisen})) as  functional dependence for the NN
and NNN exchange constants. As input parameters we use the bulk
exchange constants $J_{1,0}^{\text{MC}} $ and $J_{2,0}^{\text{MC}}
$ determined in the previous section and treat the power law
exponents $n_1$ and $n_2$ in Eqn.~(\ref{eqn:grueneisen}) as
adjustable parameters. It turns out, that because the NN exchange
$J_1$ in EuO and EuS is always much larger than the NNN exchange
$J_2$, the $T_C(a)$ dependence is quite insensitive to the
variation of $J_2$ as a function of lattice constant, i.e., the
fit yields only unreliable values for $n_2$. Because for EuTe we
have already confirmed Bloch's 10/3 law for the volume dependence
of the NNN superexchange integral $J_2$, we have therefore chosen
to fix the distance dependence of $J_2(a)$ proportional to
$r_2^{-10}$ for EuO and EuS as well. From the fit, we then obtain
$n_1^{\text{EuO}} = 19.6\pm0.4$ and $n_1^{\text{EuS}} =
22.4\pm0.3$ as the Gr\"uneisen exponents for the NN exchange
interaction. The resulting lattice constant dependence of the
Curie temperatures $T_C(a)$ and exchange interactions $J_1(a)$ and
$J_2(a)$ are plotted in in Figs.~\ref{fig:xcfit_EuO} for EuO
and~\ref{fig:xcfit_EuS} for EuS as solid, dashed and dash-dotted
lines, respectively. Evidently, an excellent fit with the
experimental data is obtained over the whole lattice parameter
range for both materials. This is also an indication that the
choice of $n_2$ = 10 is a reasonable assumption. It is also noted
that due to the about a factor of two larger Gr\"uneisen exponent
of $J_1$ compared to that of $J_2$, at high hydrostatic pressures
(small lattice constants), the NN exchange $J_1$ is as much as
five times larger than the NNN exchange $J_2$. Thus, the
ferromagnetic NN exchange completely dominates the magnetic
behavior of both materials.

\section{Exchange interaction in E\lowercase{u}S\lowercase{e}\label{sec:euse}}

\begin{figure}
 \includegraphics{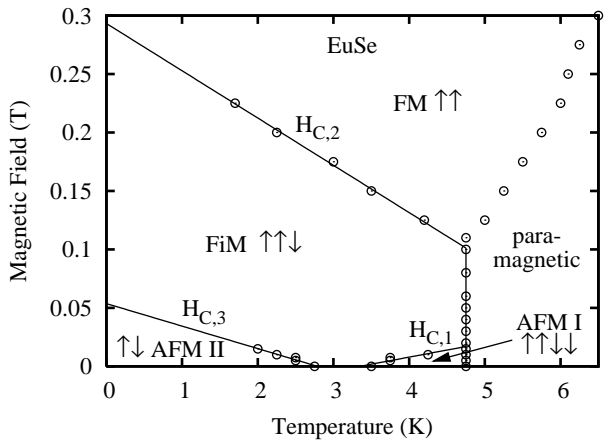}
 \caption{Experimental magnetic phase diagram of EuSe. The
phase boundaries (solid lines) are obtained
from a linear least square fit to experimental data (open symbols). See
Ref.~\onlinecite{lechn05} for details.
 \label{fig:unstrained_HTc}}
\end{figure}

\begin{figure}
  \includegraphics{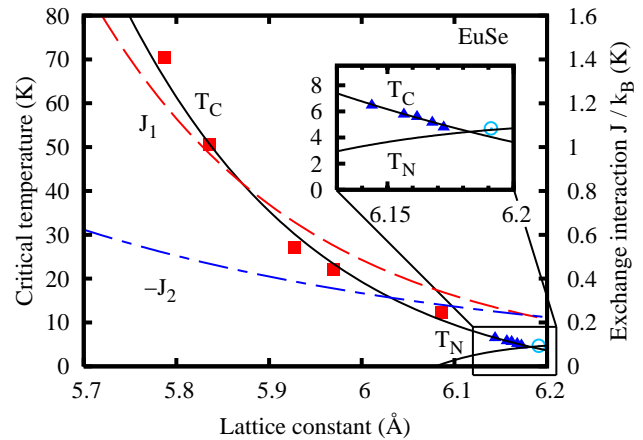}%
  \caption{(color online) Critical temperatures and exchange
interactions in EuSe as functions of the lattice parameter;
filled symbols---triangles\cite{fujiw82} and squares\cite{gonch98} correspond to
experimental Curie points. The solid line ($T_C$) represents
the least square fit of the Monte Carlo $T_C(J_1,J_2)$ to the experimental
Curie points based on the magnetic Gr\"uneisen law $J_i(r_i) \sim
r_i^{-n_i}$. The corresponding nearest and next-nearest neighbor
exchange interaction, $J_1$ and $J_2$ as
functions of the lattice parameter $a$, are represented as the dashed
and dash-dotted lines, respectively. The inset shows the
AFM~I--FM transition occurring in EuSe at hydrostatic pressures around 0.5\,GPa.
The solid graph ($T_N$), which is based on an extrapolation of $J_1(a)$ and
$J_2(a)$ towards $a_0$ corresponds to a theoretical AFM~II N\'eel temperature
which is energetically compareable to the experimentally observed AFM~I N\'eel
point (open symbol\cite{fujiw82,lechn05}). \label{fig:xcfit_EuSe}}
\end{figure}

Unlike EuO, EuS and EuTe, which exhibit stable magnetic low
temperature phases, EuSe is a metamagnet with at least four
different known ordered magnetic phases, i.e., two
antiferromagnetic phases of type~I (AFM~I) and type~II (AFM~II), a
ferrimagnetic phase (FiM) and a ferromagnetic phase (FM).
Figure~\ref{fig:unstrained_HTc} shows the $H$-$T$ phase diagram of
unstrained EuSe derived from susceptibility measurements on
several micrometer thick epitaxial layers\cite{lechn05}, and the
corresponding different spin configurations are illustrated by the
arrows. The phase boundaries shown in
Fig.~\ref{fig:unstrained_HTc} are in good agreement with earlier
publications\cite{gries71,fukum85}. In particular, the AFM~I
N\'eel point of $T_{N,\text{AFM~I}} = 4.7$\,K and the critical
field of $H_{C,3}(0) = 0.05$\,T for the transition from AFM~II to
FiM obtained by linear extrapolation of the experimental AFM~II to
FiM phase boundary are in excellent agreement with
Refs.~\onlinecite{gries71} and~\onlinecite{fukum85}.

The metamagnetic behavior of EuSe at ambient pressure shows
similarities to the situation in EuTe at the AFM~II to FM
transition that occurs at a hydrostatic pressure of 9\,GPa (see
Fig.~\ref{fig:xcfit_EuTe}), where $J_1 \approx |J_2|$. The appearance of an
AFM~I
phase at this pressure that accompanies the AFM~II to FM phase
transition\cite{gonch97} shows that at this pressure EuTe is
metamagnetic too. Thus, in both materials a metamagnetic behavior
occurs when $J_1$ and $|J_2|$ are approximately equal and cancel
each other. Then otherwise negligible additional interactions come
into play. As shown by Fig.~\ref{fig:xcfit_EuSe}, when applying hydrostatic
pressure
of 0.5\,GPa\cite{fujiw82} EuSe becomes a stable ferromagnet and at
increasing pressures up to 15\,GPa, i.e., 6\,\% reduction of the
lattice constant, the ferromagnetic ordering temperature increases
from 4.7 to above 70\,K\cite{gonch98}. The corresponding
experimental $T_C(a)$ data of EuSe is displayed as filled symbols
in Fig.~\ref{fig:xcfit_EuSe}. Lechner et al.\cite{lechn05} also
showed that the introduction of only little biaxial strain in EuSe
drastically expands the boundaries of the AFM~II phase and causes
the AFM~I phase to disappear completely. Thus, an AFM~II to
paramagnetic N\'eel point $T_{N,\text{AFM~II}}$ is observed in
strained EuSe, which is again similar to the situation in EuTe.

The magnetic phase diagram of EuSe and its
metastability for already small lattice deformations\cite{lechn05}
cannot be described by isotropic NN and NNN Heisenberg exchange
interactions alone. Especially, the ferrimagnetic and the AFM~I
phases, which show a magnetic structure with a periodicity of
three, respectively, four atomic layers require further distant
exchange interactions and/or other types of magnetic interactions
such as long-range dipolar interactions\cite{fukum85}. Moreover,
using MFA and a Hamiltonian, which includes the exchange
interaction up to the third nearest neighbor ($J_3$) and dipolar
interactions Fukuma et al.\cite{fukum85} showed that the critical
field $H_{C,3}$ at $T=0$ is independent of $J_3$ and depends only
on the sum $(J_{1}+J_{2})$ as well as the dipolar coupling
strength $D$. A good estimate of $(J_{1}+J_{2})$ can be calculated
using the mean field relation
\begin{widetext}
\begin{equation}\label{eqn:j1pj2}
 2(J_{1,0}+J_{2,0})=-\frac{1}{3\,S}g\,\mu_B\,H_{C,3}(0)-\left[
D_{xx}(\bm{Q_L})-\frac{1}
{9}D_{xx}(\bm{0})-\frac{8}
{9}D_{xx}\left(\frac{2}{3}\bm{Q_L}\right) \right]\quad.
\end{equation}
\end{widetext}
Here $D_{xx}(\bm{Q_L})$, $D_{xx}(\bm{0})$ and $D_{xx}(2\,\bm{Q_L}/3)$
correspond to the dipole coupling strength for spins
lying in the (111) plane (see Ref.~\onlinecite{fukum85} for exact definitions).
Inserting $H_{C,3}(0) = 0.05$\,T and the calculated values of $D_xx$ given
in Ref.~\onlinecite{fukum85} in Eqn.~(\ref{eqn:j1pj2}), we obtained
$(J_{1 }+J_{2})/k_B=-5.2$\,mK. Unlike for the other Eu$X$ compounds,
we were not able to derive any further reliable condition that would
allow to determine $J_{1}$ and $J_{2}$ independently of the high
pressure data in EuSe.

\begin{figure*}
  \includegraphics{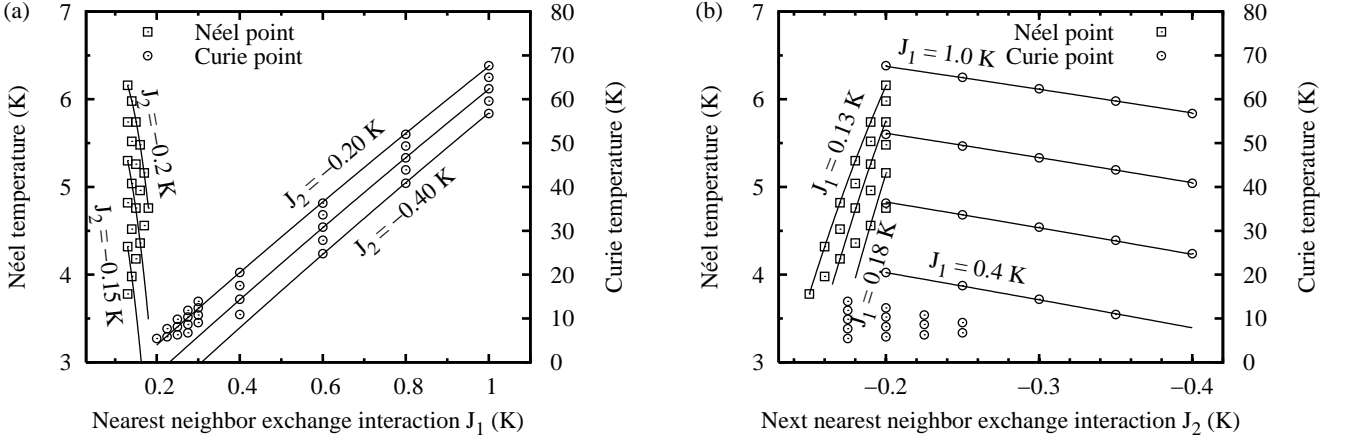}%
  \caption{Monte Carlo calculated critical temperatures
of a bulk \textit{fcc} Heisenberg system as a function of the strength of (a)
the nearest and (b) next nearest neighbor exchange interactions $J_1$ and
$J_2$, respectively, showing ferromagnetic to paramagnetic
transitions if $J_1>|J_2|$ and antiferromagnetic to paramagnetic transitions if
$J_1<|J_2|$. The parameter range corresponds to EuSe
under hydrostatic pressures between 0 and around 15\,GPa. Solid lines represent
polynomial fit functions, which are quadratic in $J_1$ and $J_2$.
\label{fig:euse_bulk_TcJi}}
\end{figure*}

Since our Heisenberg spin model can only generate FM and AFM~II orderings,
we are only able to reproduce the behavior of $T_C$ as a function of
 lattice constant in the strain-induced ferromagnetic phase of EuSe
at pressures above 0.5\,GPa. In this region, third nearest
neighbor exchange and dipolar interactions are not expected to
contribute significantly to the ferromagnetic ordering. Therefore,
we calculated the ferromagnetic ordering temperature as functions
of the NN and NNN Heisenberg exchange interaction in the range
$0.2\,\mbox{K} \leq J_1 \leq 1.0\,\mbox{K}$ and $-0.175\,\mbox{K}
\geq J_2 \geq -0.4\,\mbox{K}$ with $J_1 > |J_2|$. As shown in
Fig.~\ref{fig:euse_bulk_TcJi}, $T_C(J_1,J_2)$ is slightly
nonlinear in both $J_1$ and $J_2$, but approaches the relation
given in Eqn.~(\ref{eqn:mc_curie_temp}) for $J_1 \gg -J_2$. We
also simulated the AFM~II to paramagnetic transitions for $J_1 <
|J_2|$ in the interval $0.13\,\mbox{K} \leq J_1 \leq
0.18\,\mbox{K}$ and $-0.15\,\mbox{K} \geq J_2 \geq
-0.2\,\mbox{K}$. The calculated critical temperature $T_N$ as a
function of the exchange constants $J_1$ and $J_2$ are shown in
Figs.~\ref{fig:euse_bulk_TcJi}(a) and~(b) as open squares. As in
the case of EuTe, $T_N(J_1, J_2)$ is strongly nonlinear and both
$T_C(J_1,J_2)$ and $T_N(J_1, J_2)$ were approximated by second
order polynomials as given in Eqn.~(\ref{eqn:poly_biquad}).

Inserting the magnetic Gr\"uneisen law of
Eqn.~(\ref{eqn:grueneisen}) $J_1(r_1)=J_1(a/\sqrt{2})$ and
$J_2(r_2)=J_2(a)$ into the obtained $T_C(J_1,J_2)$ dependence, the
distance dependence of the EuSe exchange constants $J_i(a)$ was
again obtained by fitting the calculated $T_C(a)$ to the
experimental Curie points of EuSe under hydrostatic pressure
(filled symbols). Other than in the preceding sections, not only
the Gr\"uneisen exponents $n_1$ and $n_2$ but also the ambient
pressure exchange constants $J_{1,0}$ and $J_{2,0}$ were used as
adjustable parameters in the fit routine, only restricted by the
condition $(J_{1,0}+J_{2,0})/k_B=-5.2$\,mK, obtained from the
critical field $H_{C,3}(0)$ as described above. Unlike the
situation in EuO and EuS, the bahavior of $J_2(a)$ influences the
magnetic ordering considerably in the region close to $a_0$.
Eventually, $J_{1,0}/k_B = 0.223\pm0.016$\,K, $J_{2,0}/k_B =
-0.228\pm0.016$\,K, $n_1 = 24.9\pm1.8$ and $n_2 = 12.2\pm6.0$ are
obtained by the fit. As is demonstrated by the solid line in
Fig.~\ref{fig:xcfit_EuSe}, with these parameters the experimental
$T_C(a)$ data are precisely reproduced. The dependence of $J_1$
and $J_2$ versus lattice constant are depicted as dashed,
respectively, dash-dotted lines in Fig.~\ref{fig:xcfit_EuSe}. At
the bulk EuSe lattice constant of $a_0$ = 6.191 \AA, the
calculated $J_2$ is slightly larger in absolute value than $J_1$.
This changes drastically as the lattice constant is reduced under
hydrostatic pressure, with $J_1$ crossing $J_2$ already at low
hydrostatic strain and $J_1$ becoming the dominant exchange
interaction for $a < 6.15$\,\AA. The Gr\"uneisen exponents for the
NN and the NNN exchange interaction are again in good agreement
with the results obtained for EuTe, EuO and EuS.

To further justify our results on EuSe, we substituted the
obtained $J_1(a)$ and $J_2(a)$ into the theoretical $T_N(J_1,
J_2)$ AFM~II to paramagnetic N\'eel function, obtained from the
fit of Eqn.~(\ref{eqn:poly_biquad}) to the squares in
Fig.~\ref{fig:euse_bulk_TcJi}. Extrapolating $T_N(J_1(a), J_2(a))$
to the bulk lattice constant $a_0=6.191$\,\AA\ of EuSe, we
obtained $T_{N,\text{AFM~II}}^{\text{MC}}(a_0) = 4.5$\,K, which
is, as expected, above the experimentally observed AFM~II to FiM
transition temperature of around 2\,K\cite{gries71,lechn05}, but
below the AFM~I N\'eel point of
$4.7\pm0.1$\,K\cite{gries71,fujiw82,lechn05} (open symbol in
Fig.~\ref{fig:xcfit_EuSe}). That the calculated AFM~II to
paramagnetic transition temperature of $T_{N,
\text{AFM~II}}^{\text{MC}}(a_0)$ is very close to the
experimentally observed AFM~I to paramagnetic $T_{N,\text{AFM~I}}$
is also expected in mean field theory, where
\begin{eqnarray}
 T_{N,\text{AFM~I}}^{\text{MFA}}(J_1, J_2) & = & 4\,S\,(S+1)\,J_1 \\
 T_{N,\text{AFM~II}}^{\text{MFA}}(J_1, J_2) & = & -4\,S\,(S+1)\,J_2\quad.
\end{eqnarray}
Thus, $T_{N,\text{AFM~I}}$ and $T_{N,\text{AFM~II}}$ are nearly
equal when $J_1$ and $J_2$ are almost equal in strength. Moreover,
it can easily be shown, that taking third-nearest neighbor or
biquadratic exchange terms into account would not favor either of
the two antiferromagnetic ordering types (at least in the
mean field approximation) and the influence of dipolar
interactions on the ordering temperatures is typically of the
order of less than one Kelvin (see, e.g., chap.~4 of
Ref.~\onlinecite{blund01}). In addition, Lechner et
al.'s\cite{lechn05} results on biaxially strained EuSe indicate,
that the AFM~I ordering observed in unstrained EuSe with
$T_{N,\text{AFM~I}} \approx 4.7$\,K is energetically only slightly
lower than the N\'eel point of the AFM~II ordering and small
biaxial strain produces a transition from one to the other.

\section{Discussion\label{sec:discussion}}

\begin{table}
  \begin{ruledtabular}
    \input{Figs/tabular.tex}
  \end{ruledtabular}
\caption{Distance dependence of nearest (subscript 1) and next-nearest
neighbor (subscript 2) exchange interactions of europium chalcogenides under
hydrostatic pressure as obtained in our Monte Carlo study. For hydrostatic
strains of typically less than $\approx 10$\,\%, the exchange interactions can
equally well be described by either magnetic Gr\"uneisen
laws, $J_i=J_{i,0}(r_i/r_{0,i})^{-n_i}$ or by simple exponential laws,
$J_i=J_{i,0}\exp[-\alpha_i\,(r_i-r_{0,i})]$---values labelled by $^{\ast
)}$ are assumptions. \label{tab:distance_dependencies}}
\end{table}

\begin{figure}
  \includegraphics{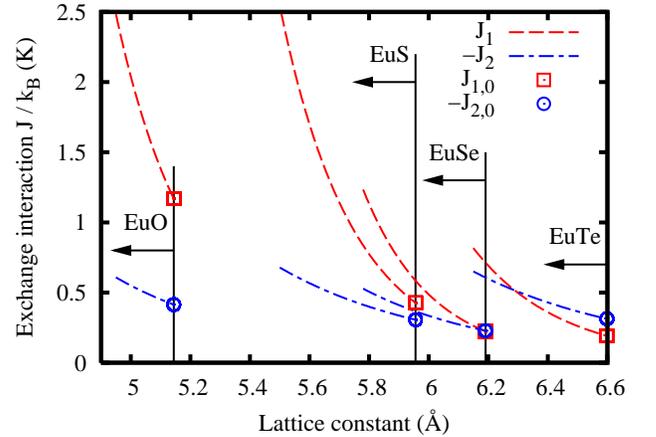}%
  \caption{(color online) Exchange
interactions in europium chalcogenides as functions of the lattice parameter.
\label{fig:xcfit_EuX_combi}}
\end{figure}

The exchange constants of all four Eu$X$ compounds and their
dependence on the lattice parameter obtained by our MC analysis
are summarized in Tab.~\ref{tab:distance_dependencies} and
Fig.~\ref{fig:xcfit_EuX_combi}. Evidently, in all cases, the
 bulk exchange constants $J_{1,0}$ and $J_{2,0}$ are almost
a factor of two larger compared to the values reported in previous
works, which in their analysis did not take spin fluctuations into
account. More importantly, we have found that the magnetic
properties and critical phase transition temperatures of all Eu$X$
compounds as a function of hydrostatic strain, i.e., lattice
parameter, can be consistently described by the magnetic
Gr\"uneisen power law dependence $J_i(r_i) = J_{i,0}
(r_i/r_{i,0})^{-n_i}$ with characteristic power law exponents of
$n_1 \approx 10$ for the NN exchange $J_1$ and of $n_2 \approx 20$
for the NNN exchange $J_2$ as indicated in
Tab.~\ref{tab:distance_dependencies}.

The corresponding dependence of the exchange interactions on the
lattice constant $J_1(a)$ and $J_2(a)$ for all compounds are
represented as dashed lines in Fig.~\ref{fig:xcfit_EuX_combi}.
Evidently, neither $J_1(a)$ nor $J_2(a)$ are continuous functions
over different members of the Eu$X$ family. Therefore, the
effect of the substitution of the anion elements cannot be
simplified to a variation in the lattice constant alone, as was
already noted in the previous work of Goncharenko et al.\cite{gonch00}.
In fact, as shown by Fig.~\ref{fig:xcfit_EuX_combi}, for a fixed lattice
constant the
absolute values of the exchange constants $J_1$ and $J_2$ are
always much larger for the compound with larger anion element,
i.e.,
\begin{equation}
 |J_i^{\text{EuTe}}(a)| \gg |J_i^{\text{EuSe}}(a)| \gg
|J_i^{\text{EuS}}(a)| \gg  |J_i^{\text{EuO}}(a)|
\end{equation}
For this reason, at a given lattice parameter the ordering
temperature is always significantly larger in EuTe compared to
EuSe, EuS and EuO. This is consistent with the extrapolation of
the $T_C(a)$ data obtained from experiments.

Apart from the empirical magnetic Gr\"uneisen law of
Eqn.~(\ref{eqn:grueneisen}) as the inter-ion distance dependence
of the exchange interactions in the Eu$X$s, we have also explored,
if different functional dependencies might reproduce the
experimental ordering temperatures as functions of the lattice
constant as well. Due to the strong superlinear increase of the
ferromagnetic ordering temperature with decreasing lattice
constant and the fact that $T_C(J_1,J_2)$ behaves very close to
linear in all of the Eu$X$ compounds, a linear distance dependence
of the exchange interactions can be categorically ruled out. One
other empirical form of $J(r)$, which could be expected from the
nature of quantum mechanical two-electron multicenter integrals,
e.g., the Heitler-London approach to solve the Schr\"odinger
equation for the hydrogen molecule\cite{heitl27}, is a simple
exponential distance law of the form
\begin{eqnarray}
 J(r) & = & J_0\exp\left[-\alpha\,(r-r_0)\right]
              \label{eqn:simple_exponential} \\
      & = & J_0\exp\left[-\alpha\,r_0\,(r/r_0-1)\right]
              \label{eqn:simple_exponential_rel}\quad,
\end{eqnarray}
where $J(r_0) = J_0$ is again the exchange constant under normal
condition. Applying this exponential law in the simulated critical
temperatures $T_N(J_1,J_2)$ and $T_C(J_1,J_2)$ and fitting the
experimental critical points of the Eu$X$s as functions of the
lattice constant, it turns out, that the results for $J_1(a)$ and
$J_2(a)$ are practically \emph{indistinguishable} from those
received by exploring the Gr\"uneisen power law. In
essence, the fitted $T_N(a)$ and $T_C(a)$ curves and the $J_1(a)$
and $J_2(a)$ distance dependencies of the exchange interactions
coincide almost exactly with those presented in Fig.~\ref{fig:xcfit_EuX_combi}.
The reason for this unambiguity is that the
variations in the interatomic distances achievable by hydrostatic
strain are generally too small ($\leq 8$\,\%) to be able to
definitely single out between the two functional dependencies.
However, while the
distance dependencies of the exchange
interaction for the different members of the Eu$X$ family show
consistant Gr\"uneisen exponents $n_i$ for the NN and the NNN
exchange interactions, such similarities can not be expected for
the scaling factors $\alpha_i$ in the simple exponential
description. This follows from the fact that only the magnetic
Gr\"uneisen law of Eqn.~(\ref{eqn:grueneisen}) is defined in terms
of relative changes of the lattice parameter $r/r_0$. This, however, can also
be achieved by redefining the exponential distance law as a function of $r/r_0$
(Eqn.~(\ref{eqn:simple_exponential_rel})). Then $\alpha\,r_0$ represents a
dimensionless scaling factor that can be compared for the different members of
the Eu$X$ family. The calculated scaling factors $\alpha_i$ are given together
with the $n_i$ and the $J_{0,i}$ in Tab.~\ref{tab:distance_dependencies} and are
in the range of 4.5--6\,\AA$^{-1}$ for $\alpha_1$ and 1.6--2.3\,\AA$^{-1}$ for
$\alpha_2$. As it turns out, the values for $\alpha_i\,r_0$ are approximately
equal to the respective $n_i$s for all Eu$X$ compounds. This becomes clear from
a more mathematical point of view. Using the relation\cite{coffm79}
\begin{equation}
  \alpha = -\frac{\frac{dJ}{dr}(r)}{J(r)}\quad,
\end{equation}
and requiring that both $J(r)$ and its derivative with respect to $r$ have to
match at $r_0$ yields the relation $n = \alpha\,r_0$. Thus, we have shown
that for small deviations of $r$ from $r_0$ the power law and the exponential
dependence are nearly equivalent when $\alpha$ is chosen as $\alpha = n/r_0$.

The theoretical derivation of analytic scaling
laws for the distance dependence of the exchange interactions is
far from trivial and estimates can only be obtained as far as the
exchange mechanisms are understood. According to
Kasuya\cite{kasuy70} the most important contribution to the NN
exchange interaction consists of a virtual excitation of a
Eu\textsuperscript{2+} 4\textit{f} electron to the 5\textit{d}
state of a NN cation and a subsequent intra-atomic
\textit{d}-\textit{f} exchange. For this type of exchange, a
distance dependence $J_1(r_1) \sim \exp(-8\,r_1/r_0)$ is
considered. The NNN exchange may consist of several competing
components, which are considered to involve excitations of the
anion \textit{p} electrons to neighboring cation 5\textit{d}
states. To our knowledge Kasuya did not give an estimate for the
distance dependence of the NNN superexchange. Lee and
Liu\cite{lee84} on the other hand proposed interband exchange
mechanisms for both $J_1$ and $J_2$ where the exchange of the
localized 4\textit{f} moments is mediated by virtual excitations
of chalcogenide-valence band \textit{p} electrons into the empty
Eu\textsuperscript{2+} 5\textit{d} conduction bands, together with
a subsequent interband exchange of the \textit{d} electron
(\textit{p} hole) with the localized 4\textit{f} electrons. In
this semiconductor analogue of the Ruderman-Kittel-Kasuya-Yosida
(RKKY) interaction $J(r) \sim r^{-4}$ is considered for the
distance dependencies of the NN and NNN exchange interactions.
Both estimates for the distance dependence of the exchange
interactions more or less support the
picture\cite{mcgui64,zinn76,wacht79,mauge86} of the exchange
interactions being continuous functions across different members
of the Eu$X$ family. Our analysis clearly shows that the exchange
interactions are varying much stronger with the interatomic
distances than previously assumed and that the exchange of the
anion element can not be attributed solely to shifts in the
lattice parameter. With a Gr\"uneisen exponent of $n_1 \approx
20$, there is a particularly strong dependence of the NN exchange
$J_1$ as function of the interatomic distances and the exponent
$n_2 \approx 10$ for the distance dependence of the NNN exchange
interaction $J_2$ coincides with Bloch's\cite{bloch66} empirical
law for the volume dependence of the superexchange in magnetic
solids.

The possibility to integrate EuO with Si and
GaN\cite{letti03,schme07} together with the fact that $T_C$ can be
increased by doping\cite{konno96,ott06} and hydrostatic
strain\cite{dimar87,moser88,abdel90} to temperatures up to 200\,K
led to a renewed interest in the ferromagnetic europium
monochalcogenides as possible materials for future spintronic
devices. As hydrostatic pressure is not an option for practical
applications, epitaxial strain has been suggested as an
alternative way to increase the ferromagnetic ordering temperature
in EuO\cite{ingle08}. Ingle and Elfimov showed in their ab initio
study that biaxial compressive strain increases $T_C$ in EuO
similarly to the situation in EuSe\cite{lechn05} and
EuTe\cite{kepa03,schie08}, where the antiferromagnetic ordering
temperature is increased with a reduction in the in-plane lattice
constant of ultrathin (111) oriented epilayers. As shown by
Ref.~\onlinecite{schie08}, by proper adjustment of the exchange
constants, the behavior of epitaxially strained EuTe layers can
well be described by our Monte Carlo calculations.

\section{Conclusion}
\label{sec:conclusion}

In summary, we have applied the Monte Carlo method to determine
the exchange integrals in the Eu$X$ compounds. To this end, we
have determined the general dependences of the magnetic ordering
temperatures, i.e., ferromagnetic $T_C(J_1,J_2)$ and
antiferromagnetic $T_N(J_1,J_2)$ as functions of the  NN and NNN
exchange interactions of a system of classical Heisenberg spins at
the sites of an \textit{fcc} lattice. This was subsequently
applied to determine the distance dependence of the NN and NNN
exchange interactions from hydrostatic pressure experiments based
on the magnetic Gr\"uneisen law\cite{bloch66} $J_i(r_i) \sim
r_i^{-n_i}$, where $r_i$ denotes the interatomic distances between
neighboring Eu ions. It turns out that the distance dependences of
the exchange interactions $J_1(r_1)$ and $J_2(r_2)$ of the
different members of the Eu$X$ family can be consistently
described by Gr\"uneisen exponents $n_1\approx 20$ and $n_2\approx
10$, where the latter conforms with Bloch's empirical 10/3 law for
the volume dependence of the superexchange
interaction\cite{bloch66}. The strong dependence of the exchange
constants on the lattice parameter provides room for substantially
increasing the magnetic ordering temperatures in strained
heteroepitaxial structures, which is an important prerequisite for
device applications.















\begin{acknowledgments}
This work was supported by the Austrian Science funds FWF
and GMe and the Austrian NANO Initiative (NSI). The authors would like to thank
Reinhard Folk for usefull hints
and valuable disscussions and Daniel Gruber and Johann Messner
for technical assistance.

\end{acknowledgments}

\bibliography{Bib/eux_distance_laws}

\end{document}

%% file: Figs/tabular.tex
\begin{tabular}{lcccc}
                       & EuTe  & EuSe  & EuS            & EuO   \\
\hline
$a_0$ (\AA)            & 6.598 & 6.191 & 5.956          & 5.144 \\
$T_N(a_0)$ (K)         & 9.85  & 4.7   &  -             &  -    \\
$T_C(a_0)$ (K)         &  -    &  -    &16.6            &69.15  \\
\hline
$J_{1,0}$ (K)          & 0.192 & 0.223 & 0.427          & 1.169 \\
$n_1$                  &20.6   &24.9   &22.4            &19.6   \\
$\alpha_1$ (\AA$^{-1}$)& 4.56  & 5.99  & 5.47           & 5.44  \\
\hline
$J_{2,0}$ (K)          &-0.313 &-0.228 &-0.306          &-0.414 \\
$n_2$                  &10.4   &12.2   &10$^{\ast )}$ &10$^{\ast )}$ \\
$\alpha_2$ (\AA$^{-1}$)& 1.63  & 2.31  & 1.68$^{\ast )}$& 1.94$^{\ast )}$
\end{tabular}